\documentclass[journal=JCTC,manuscript=letter,layout=twocolumn]{achemso} 
\usepackage[english]{babel}
\usepackage[utf8]{inputenc} 
\usepackage[table]{xcolor} 
\usepackage[%
  colorlinks=true,
  urlcolor=blue,
  linkcolor=blue,
  citecolor=blue
]{hyperref}
\usepackage{algorithm} 
\usepackage{algpseudocode} 
\usepackage{bbold} 
\usepackage{amssymb}
\usepackage{bm}
\usepackage{bbm}
\usepackage{tabularx, booktabs}
\usepackage{xspace}
\usepackage{listings}
\usepackage{lipsum}
\usepackage{chemformula} 
\usepackage{physics} 
 
\newcolumntype{Y}{>{\centering\arraybackslash}X}
\definecolor{dgreen}{rgb}{0,.5,0}
\definecolor{dblue}{rgb}{0,0,.5}
\definecolor{dred}{rgb}{0.5,0,.5}

\usepackage{soul}
 
\title{Orthogonally Constrained Orbital Optimization: \\ assessing changes of optimal orbitals for orthogonal multi-reference states} 

\author{Saad Yalouz}
\email{yalouzsaad@gmail.com}
\affiliation{Laboratoire de Chimie Quantique, Institut de Chimie,
CNRS/Université de Strasbourg, 4 rue Blaise Pascal, 67000 Strasbourg, France}

 \author{Vincent Robert} 
\affiliation{Laboratoire de Chimie Quantique, Institut de Chimie,
CNRS/Université de Strasbourg, 4 rue Blaise Pascal, 67000 Strasbourg, France}

\begin{document}

\begin{abstract}\def\ddroit{{\rm d}} 
The choice of molecular orbitals
is decisive in configuration interaction calculations. In this letter,  
a democratic description of the ground and excited states
follows an orthogonally constrained orbitals optimization 
to produce state-specific orbitals. 
The approach faithfully recovers the excitation energy of a
four-electron Hubbard trimer, 
whereas state-average calculations
can miss the value by a factor 2.5.
The method emphasises the need for orbitals 
optimization
to reduce expansions and to reach spectroscopic accuracy.

\end{abstract}

\maketitle

\section{Introduction}

Electron transfer reactions are determinant in many different fields, ranging from biophysical processes to artificial 
compounds with potential applications.
Ever since the synthesis of the emblematic Creutz-Taube compound~\cite{creutz1969direct}, molecular chemistry has produced a wealth of compounds, combining different redox 
centers and linkers. In this context, mixed-valence inorganic compounds are excellent model systems to investigate such
phenomenon. The valences can  be  either trapped, interconvertible with a low activation barrier, or completely delocalized. 
An intense intervalence charge transfer band characterizes the last categories, and magnetic exchange couplings
can be observed.  

Accurate methods of investigation are desirable to accurately determine band shapes and transition energies. 
Regardless their computational costs, the objectives 
are two-fold.  
First, spectroscopic accuracy is a prerequisite to validate their robustness
and deliver means of interpretations of experimental observations. Then, the microscopic
information available in the ground and excited states is a valuable contribution
to rationalize the leading phenomena and to construct model Hamiltonians.
From  variational methods, the ground state $\ket{\Psi_0}$ can be determined using different
strategies (wavefunction theory (WFT) or density functional theory (DFT)). 
The construction of excited states $\ket{\Psi_I}$ is much more problematic, in particular 
when $\ket{\Psi_0}$ and $\ket{\Psi_I}$
share the same spin and space symmetries. 
Even though stationary points of the expectation value of the Hamiltonian give access to excited states
(Ritz's theorem), the strategy and its implementation are not straightforward.
As a major breakthrough, the maximum overlap method was designed
to converge on higher solutions
of the self-consistent field (SCF) equation
\cite{gilbert2008self,barca2018simple}.
Despite the loss of orthogonality between the SCF 
solutions, the method has produced a wealth of excitation energies in different compounds~\cite{gilbert2008self}.
The key role of the molecular orbitals (MOs) in describing electron
transfer processes was also stressed for the intervalence
charge transfer of a synthetic nonheme binuclear
mixed-valence compound~\cite{domingo2015electronic}.

More recently, selected configuration interaction (CI) calculations
(\textit{e.g.} Configuration Interaction using a Perturbative Selection made Iteratively, CIPSI) 
and quantum Monte Carlo simulations~\cite{dash2021tailoring,cuzzocrea2022reference}
were performed based on the
``largest technically-affordable number of determinants'' constructed on common set
of natural orbitals (\textit{i.e.} state-average). 
Excellent agreement with high-level coupled cluster references
was reached in the determination of excitation energies.
Tremendous efforts are still put into benchmarking
multi-reference excitation energies
using state-average methods and the active space selection issue~\cite{king2022large}.

\begin{figure}[!h]
    \centering
    \includegraphics[width=\columnwidth]{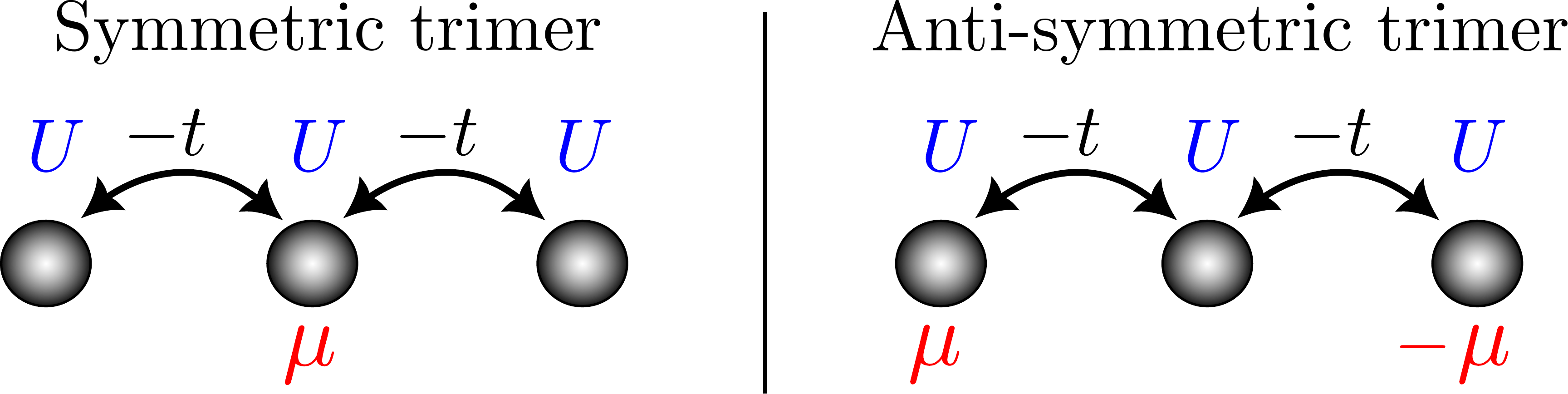}
    \caption{\textbf{Illustration of the model systems ruled by a Hubbard Hamiltonian.}} 
    \label{fig:FH_3site_STRONG}
\end{figure}

An alternative 
is the application of the variational 
method restricted to a sub-space orthogonal
to the ground state. 
Following this strategy, one would like to express 
all states on the same footing, moving away from 
the state-average strategy. 
Therefore, we designed a method here-referred to as
orthogonally constrained orbital optimization (OCOO)
to generate the first 
excited state constructed on optimized MOs
and maintaining the orthogonality with the ground state.
The method does not rely on any pre-conditioned structure
of the excited states and somewhat differ from previous
strategies~\cite{gilbert2008self,barca2018simple,hait2020excited,levi2020variational,carter2020state,gavnholt2008delta}.
Our main intention is to identify the regimes where the MOs of the excited 
state strongly differ from the 
ground state ones. The method is applied to a three-site Hubbard Hamiltonian controlled
by the on-site ($\mu$), hopping ($t$) and repulsion ($U$) energies (as illustrated in Figure~\ref{fig:FH_3site_STRONG}).

Such model is a playground to identify the limits of traditional
state-average strategies and to foresee electronic correlation 
regimes calling for different MOs basis sets.

 \begin{figure*} 
     \centering
     \includegraphics[width=\textwidth]{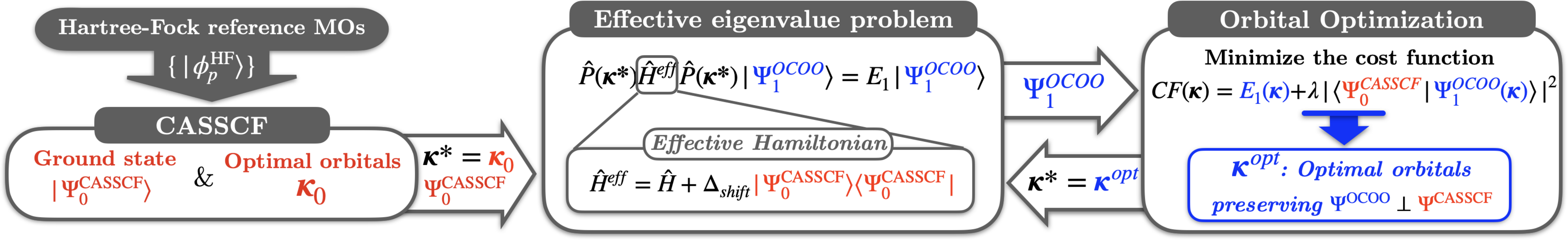}
     \caption{\textbf{Flowchart of the Orthogonally Constrained Orbital Optimization (OCOO)  algorithm.}}
     \label{fig:flow}
 \end{figure*}

\section{Orthogonally Constrained Orbital Optimization (OCOO) Method}
 In this section, we provide the details of the OCOO to describe excited states. 
 The method starts with the construction of a multi-reference
ground state in a given model active space. The
latter is constructed on a restricted number of Slater determinants 
built on orthogonal MOs.
The expansion amplitudes and
the MOs coefficients are optimized following a complete active
space self-consistent field (CASSCF) framework~\cite{siegbahn1981complete}. 
At convergence, an approximated ground state $|{\Psi}_0^\text{\tiny CASSCF}\rangle$
with energy $E_0^\text{\tiny CASSCF}$ is generated with an optimized MOs basis set $B_0$
characterized by $\boldsymbol{\kappa}_0$. Here, ``$\boldsymbol{\kappa}_0$'' relates to the way the  MOs $\ket{{\phi}_p(\boldsymbol{\kappa})}$  are parametrized during the orbital optimization process:
\begin{equation}
     \ket{{\phi}_p(\boldsymbol{\kappa})}  = \sum_l \ket{\phi_l^\text{HF}} [\mathbf{\exp}(-\boldsymbol{\kappa})]_{lp}, 
\end{equation}
$\boldsymbol{\kappa}$ being the anti-hermitian matrix generator encoding the orbital rotation parameters, and  $\lbrace \phi_l^\text{HF} \rbrace$ a set of initial MOs to be optimized (in our case, Hartree-Fock MOs). 
 
Then, the first excited state $|{{\Psi}_1^\text{\tiny OCOO}}\rangle$ is constructed  with the twofold objective that is ({\it i}) to preserve  orthogonality with the ground state (\textit{i.e.} $\langle {{\Psi}_0^\text{\tiny CASSCF}}|{{\Psi}_1^\text{\tiny OCOO}}\rangle = 0$),
and ({\it ii}) to optimize the MOs to generate a state-specific $B_1$ basis set ($\boldsymbol{\kappa} = \boldsymbol{\kappa}_1$). 
In practice, it is assumed that the multi-reference 
structure of both $| {{\Psi}_1^\text{\tiny OCOO}} \rangle $ and $|{\Psi}_0^\text{\tiny CASSCF} \rangle$ follow the same level of description.
These states are built with the same  number of ``active-space like'' Slater determinants in their respective optimized basis
sets $B_0$ and $B_1$. Therefore, any basis set modification is likely to change the physical content of the four configurations.
Let us stress that no assumption on the excited state structure is made here, in contrast
with strategies used in some reported
WFT- and DFT-based approaches (see Refs.~\cite{gavnholt2008delta,doi:10.1021/acs.jctc.8b00406,gilbert2008self}).
How do the basis $B_0$ and $B_1$ differ is at the heart of the present study.

 In practice, the first excited state energy is
estimated by solving the effective eigenvalue problem with fixed parameter $\boldsymbol{\kappa^*}$:
\begin{equation}\label{eq:eig_val_prob}
     \hat{P}( \boldsymbol{\kappa^*} )  \hat{H}^\text{eff}\hat{P}( \boldsymbol{\kappa^*} )    \ket{\Psi_1^\text{\tiny OCOO}  } = E_1   \ket{\Psi_1^\text{\tiny OCOO}  },
\end{equation}
with 
\begin{equation}\label{eq:GS_eff}
     \hat{H}^\text{eff} =  \hat{H}  + \Delta^\text{shift} |{{\Psi}_0^\text{\tiny CASSCF} }\rangle\langle{{\Psi}_0^\text{\tiny CASSCF} }|,
\end{equation}
where $\hat{H}$ is the full Hamiltonian of the system and $\hat{P}( \boldsymbol{\kappa^*} ) $ a projector over a restricted set of determinants following an active space structure in the fixed $\boldsymbol{\kappa^*}$ MOs basis set.
The effective active-space Hamiltonian $\hat{P}( \boldsymbol{\kappa^*} ) \hat{H}\hat{P}( \boldsymbol{\kappa^*} ) $  is complemented with a parametrized $\Delta^\text{shift}$  projection 
(in practice $\Delta^\text{shift} / t = 10^8$)
so that any eigenfunction with non-negligible decomposition on $|{{\Psi}_0^\text{\tiny CASSCF}}\rangle$  gets penalized.
As a result, solving Eq.~(\ref{eq:GS_eff}) produces $ \ket{\Psi_1^\text{\tiny OCOO}  } $ which is an ``effective'' ground state of the associated energy-shifted active-space Hamiltonian $\hat{P}( \boldsymbol{\kappa^*} ) \hat{H}^\text{eff} \hat{P}( \boldsymbol{\kappa^*} )$.

In the second step of the OCOO method, we act on the orbital rotation parameter $\boldsymbol{\kappa}$ to optimize the MOs. 
For this, we define a cost function to be minimized 
\begin{equation}\label{eq:CF}
    \text{CF}(\boldsymbol{\kappa}) = E_1(\boldsymbol{\kappa}) + \lambda  \left|\langle {{\Psi}_0^\text{\tiny CASSCF}} | {{\Psi}_1^\text{\tiny OCOO}(\boldsymbol{\kappa})}\rangle \right|^2.
\end{equation}
The latter includes ({\it i}) the regular orbital-rotation dependent energy 
\begin{equation} 
 E_1(\boldsymbol{\kappa}) = \bra{\Psi_1^\text{\tiny OCOO} (\boldsymbol{\kappa})  } \hat{H}\ket{\Psi_1^\text{\tiny OCOO} (\boldsymbol{\kappa})  }
\end{equation} 
(from state-specific CASSCF method), and ({\it ii}) an overlap penalty term with amplitude $\lambda$
(in practice $\lambda /t = 10^{8}$).
The role of this second contribution is to counterbalance the energy minimization with a measure of orthogonality between 
$| {{\Psi}_0^\text{\tiny CASSCF}} \rangle$
and $| {{\Psi}_1^\text{\tiny OCOO} (\boldsymbol{\kappa})} \rangle$.

A summarized flow-chart of the OCOO algorithm is given in Figure~\ref{fig:flow}.
The iterative procedure based on Eq.~(\ref{eq:GS_eff})  and Eq.~(\ref{eq:CF}) starts with $\boldsymbol{\kappa} = \boldsymbol{\kappa}_0$.
At convergence,  a multi-reference state is generated in an optimized basis set $B_1$
characterized by $\boldsymbol{\kappa} = \boldsymbol{\kappa}_1$.
Let us stress that both steps of the OCOO process call for the calculation of scalar products between multi-reference wavefunctions expressed in two different  basis sets, namely
$B_0$ and $B_1$ (as shown in Eq.~(\ref{eq:GS_eff}) and Eq.~(\ref{eq:CF})). 
Convergence  is reached  as soon as the variation  in the cost 
function Eq.~(\ref{eq:CF}) is less than $10^{-7} t$ 
between two successive iterations.

\section{Numerical results}

Keeping in mind the difficult selection of the active space, 
the present study is focused on the construction of
state-dependent MOs basis sets that preserve
the orthogonality between the multi-reference
wavefunctions.
Evidently, a two-electron in two-orbital model system is not
flexible enough to investigate orbital relaxation.
Practically, the OCOO method was implemented on a model system (four-electron three-site) 
ruled by a Hubbard Hamiltonian (see Figure~\ref{fig:FH_3site_STRONG})
inspired by a 
CAS[2,2] on top of a single inactive orbital.
Given a set of hopping $t$ and on-site repulsion $U$ energies,
the on-site potential $\mu$ value was varied and the
ground and excited states energies were calculated based
on either state-averaged CAS[2,2]SCF calculations (\textit{i.e.} a single set of MOs
to describe both states), or state-specific 
schemes (\textit{i.e.} two sets of optimized MOs, 
CASSCF and OCOO for the ground and excited states, respectively). 
Whatever the strategy, both wavefunctions were expanded on four
Slater determinants written as $\ket{1\overline{1}2\overline{2}},\ket{1\overline{1}2\overline{3}},\ket{1\overline{1}\overline{2}3}$ and $\ket{1\overline{1}3\overline{3}}$ 
where the index ``$i$'' (or ``$\overline{i}$'') stand
for the $i$th $\alpha$-MO (or $\beta$-MO).
Charge transfer phenomenon being a motivation in the present study, symmetric and anti-symmetric model systems were considered by varying the values
on-site potentials (see Figure~\ref{fig:FH_3site_STRONG}).

All numerical implementations and calculations presented in this paper were carried out within the python package \textit{QuantNBody}  (see Ref.~\cite{codeQuantNBody})  recently developed by one of us (SY). This package was designed to facilitate the numerical implementation of second quantization algebra and the manipulation of many-body wavefunctions. 
We used this numerical toolkit to build/diagonalize the Hubbard Hamiltonians, implement orbital optimizations and evaluate the non-trivial overlap between multi-reference wavefunctions expressed in different MOs basis. 
Electronic structures result from the competition between
one-electron and two-electron contributions. Therefore, 
we first derived the eigenvalues of the one-body part of the Hubbard Hamiltonians $\hat{h}$ for both systems to evaluate the so-called spectral band. 
The eigen-values are readily derived and the spectral band $\Delta \epsilon$ of the symmetric trimer reads
\begin{equation} \label{eq:spectral_band_sym}
    \Delta \epsilon =   \sqrt{\mu^2+8t^2}.
\end{equation}
For the anti-symmetric trimer, this value becomes
\begin{equation} \label{eq:spectral_band_antisym}
    \Delta \epsilon =  2\sqrt{\mu^2+2t^2}
\end{equation}

\begin{figure*} 
    \centering
    \includegraphics[width=0.95\columnwidth]{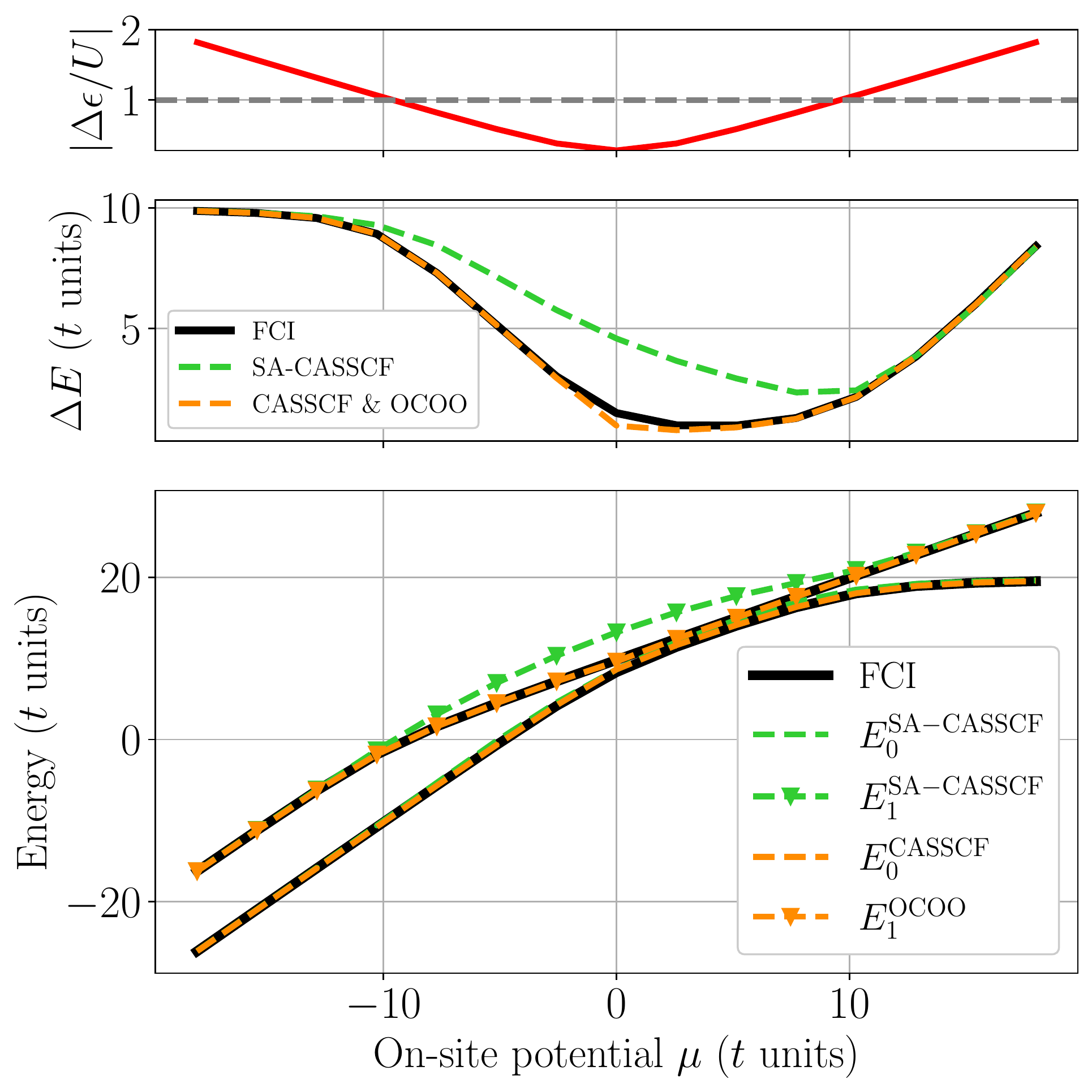}
    \includegraphics[width=0.95\columnwidth]{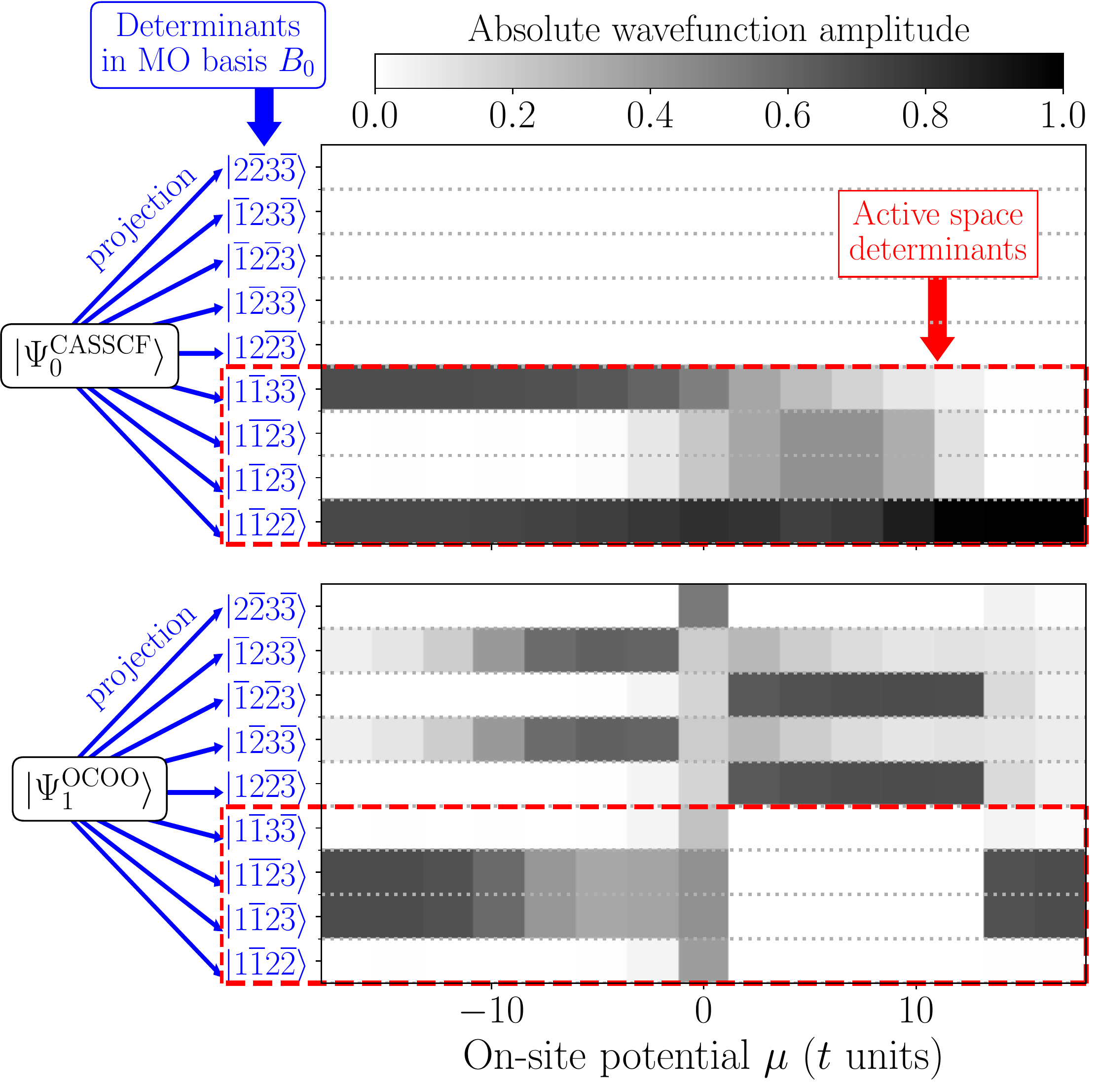}
    \caption{\textbf{Symmetric three-site Hubbard model with tunable on-site potential on the central site (strong correlation regime: $U/t=10$).}\textbf{ Left panel:} Energies  
    obtained from FCI (in black), SA-CASSCF (in green) and  with the combination CASSCF+OCOO (in orange), respectively for the ground and first excited states. The middle panel gives the vertical excitation energies obtained with the three methods. The upper panel shows the ratio $\Delta \epsilon/U$. \textbf{Right panels:}  Decompositions of the multi-reference $\ket{\Psi_0^\text{CASSCF}}$ and $\ket{\Psi_1^\text{OCOO}}$ in the $B_0$ basis set. For the latter, strong variations are observed, stressing the deep basis set modifications.}
    \label{fig:trimer_sym}
\end{figure*}

\begin{figure*} 
    \centering
    \includegraphics[width=0.95\columnwidth]{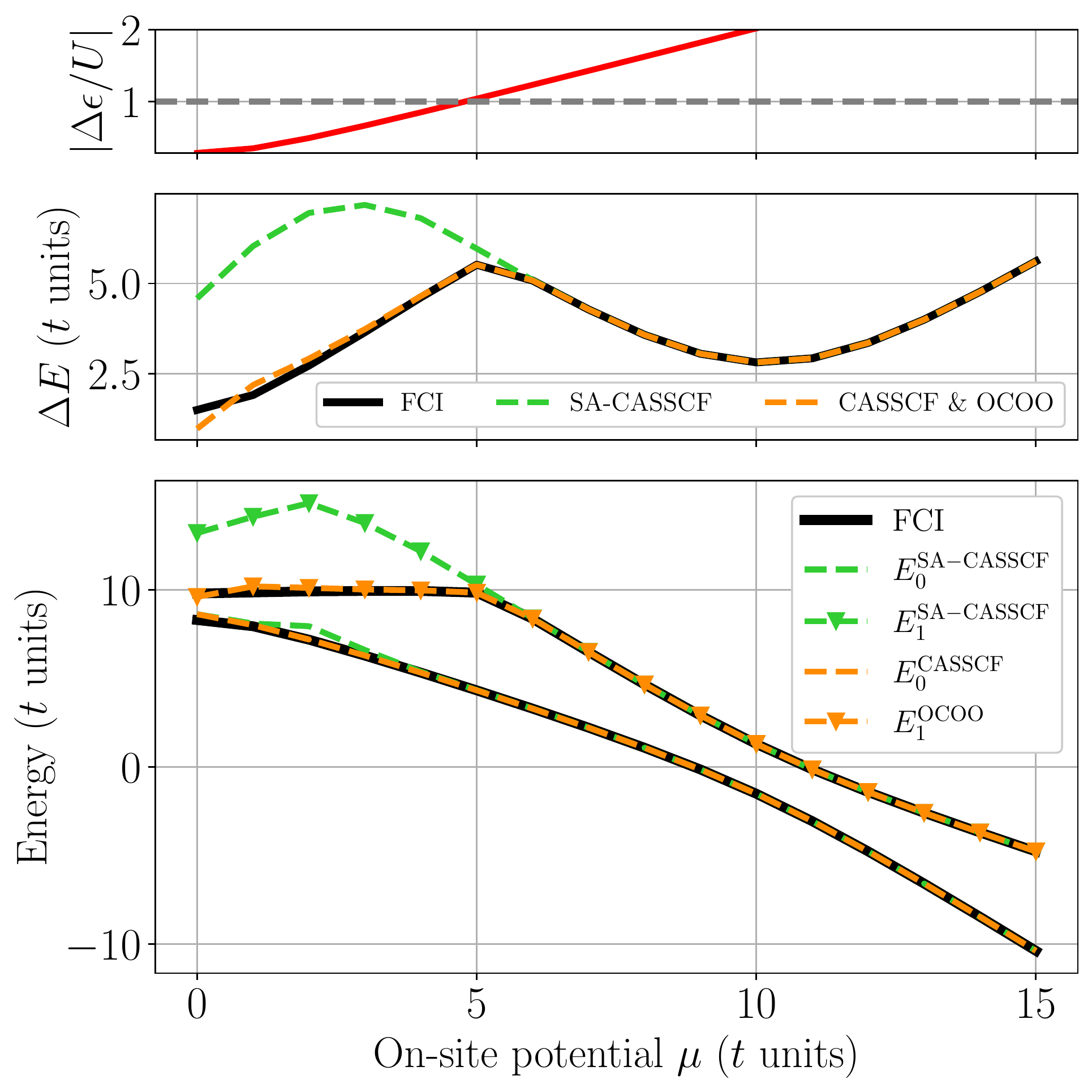}
    \includegraphics[width=0.95\columnwidth]{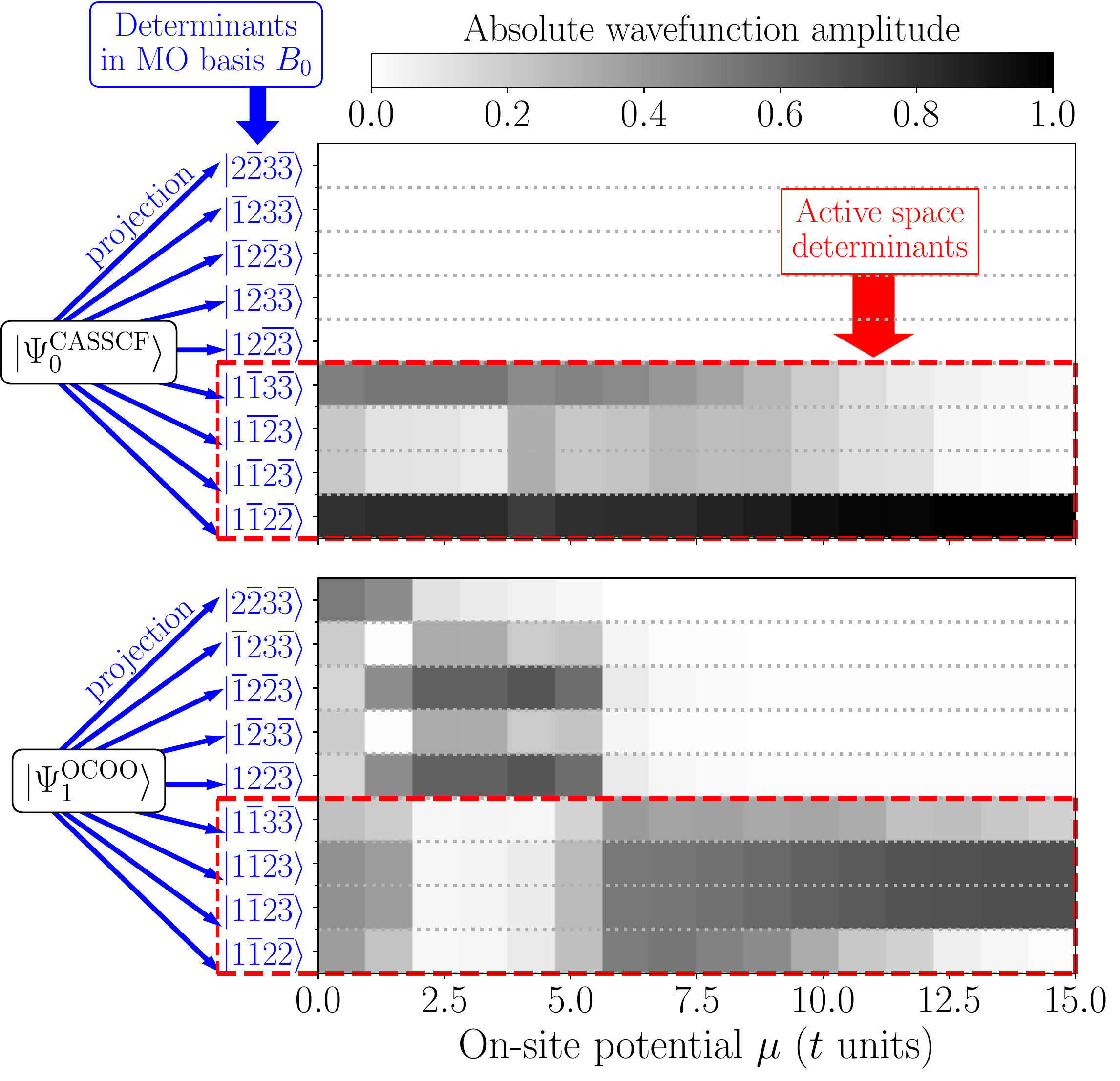}
        \caption{\textbf{Anti-symmetric three-site Hubbard model with tunable on-site potential on left and right sites (strong correlation regime: $U/t=10$).}\textbf{ Left panel:} Energies obtained with FCI (in black), SA-CASSCF (in green) and  with the combination CASSCF+OCOO (in orange) respectively for ground
        and first excited state. The middle panel gives the vertical excitation energies obtained with the three methods. The upper panel shows the evolution of the ratio $\Delta \epsilon/U$. \textbf{Right panels:} Decompositions of the multi-reference $\ket{\Psi_0^\text{CASSCF}}$ and $\ket{\Psi_1^\text{OCOO}}$ in the $B_0$ basis set. For the latter, strong variations are observed, stressing the deep basis set modifications.}
    \label{fig:trimer_asym}
\end{figure*}
Based on these model systems, the relevance of
the OCOO approach was examined and our results were 
compared to full-CI calculations in a strong
correlation regime $U/t = 10$. All results are shown 
in Figures \ref{fig:trimer_sym} and  \ref{fig:trimer_asym}.

First, as shown in the bottom left panels of the Figures, the ground state energy is faithfully reproduced by 
the CASSCF and SA-CASSCF 
calculations in the whole range of
$\mu$ values. 
From the
system size, the orbitals
relaxations are sufficient to retrieve most of the full-CI
wavefunction with 
$| \langle{{\Psi}_0^\text{\tiny CASSCF}|\Psi_0^\text{\tiny FCI}} \rangle | \sim 0.99$
for $\mu \neq 0$.
In the limit $\mu = 0$, the spectral bands are minimized (see Eq.~\ref{eq:spectral_band_sym}
and \ref{eq:spectral_band_antisym})
and the CAS[2,2] picture might be questionable (projections $\sim 0.91$).

Since the ground state energy does not
suffer from the use of SA-CASSCF MOs,
one would like to evaluate the robustness
of a state-average method in the evaluation
of vertical excitation energies noted $\Delta E$.
Strong deviations with respect to full-CI
are observed, and $\Delta E$ 
can be over-estimated by a
factor up to 2.5 (in worst case scenario seen in
Figures~\ref{fig:trimer_sym} and~\ref{fig:trimer_asym}, middle panels).
Such shortcomings of the SA-CASSCF approach are 
anticipated when  
$\Delta \epsilon / U < 1 $
as confirmed in the top left-panel of Figures~\ref{fig:trimer_sym} and \ref{fig:trimer_asym}. 
Therefore,
the SA-CASSCF excitation energy deteriorates for $\mu/t \sim 5$ (symmetric trimer) 
and $ -10 < \mu/t < 10$ (anti-symmetric trimer), revealing 
the inaccuracy of the state-average strategy.

In contrast, excellent agreement between $E_1^{\mbox{\tiny OCOO}}$ and $E_1^{\mbox{\tiny FCI}}$ is reached for both model systems
as soon as the excited state MOs are optimized following the OCOO method.
The excitation energy is perfectly recovered
as soon as each multi-reference state is individually
generated in its optimized basis set. 
A similar conclusion was reached 
from a non-systematic procedure in donor-acceptor
compounds. It was suggested that orbital relaxation should be explicitly taken into account~\cite{meyer2014charge}.

To get a better view on the MOs modifications, the excited state wavefunction 
$\ket{\Psi_1^\text{OCOO}}$ 
was expanded in the ground state basis set $B_0$.
As shown in the right panel of Figures~\ref{fig:trimer_sym} and~\ref{fig:trimer_asym}, the compact form constructed on the four configurations
is lost, a feature of a deep change between $B_0$ and $B_1$ basis sets.
Quantitatively, the projection of $\ket{\Psi_1^\text{OCOO}}$  on the CAS[2,2] subspace defined in  $B_0$  exhibits strong variations and can even become null. 
Besides, the zeroth-projection domain is reduced from $\mu/t \in \left[0, 12 \right]$ (Figures~\ref{fig:trimer_sym})
to $\mu/t \in \left[2, 3 \right]$ (Figures~\ref{fig:trimer_asym})
 upon symmetry-breaking.
These observations support the idea of the  ill-definition of the $B_0$ basis set for the compact active-space representation of the first excited state in the strong correlation regime ($U/t = 10$) of symmetrical systems (e.g. mixed-valence compounds).
Finally, note that all our conclusions and analysis remain unchanged in weaker correlation regimes (\textit{i.e.} $U/t = 5$, not shown here)
for which the domain calling for a simultaneous optimizations
of the ground and excited states orbitals is reduced.

\section{Conclusion}
An orthogonaly constrained orbital optimization (OCOO) method is suggested
and implemented to foresee the correlation regimes where orbitals relaxations
cannot be ignored. The structure of the ground state is of CASSCF-type, whereas the excited state is defined as the lowest-lying orthogonal multi-reference one.
Based on a trimer model system ruled by a Hubbard Hamiltonian, 
the excitation energy is evaluated from a multi-reference
state-specific description of orthogonal states. 
As soon as the one-site energy $U$ competes with the spectral band $\Delta \epsilon$,
SA-CASSCF calculations fail to reproduce the vertical excitation energy. 
Despite its simplicity, the model not only offers a practical 
method to democratically treat the ground and excited states, but also
identifies correlation regimes where the robustness of SA-CASSCF 
might be questionable to
reach spectroscopic accuracy. Finally, this work aims at 
re-emphasizing the importance of orbital optimization in zeroth-order multi-reference wavefunctions expansions. Explorations on the benefit of second-order perturbation treatment
are planned to stress the decisive choice of MOs basis sets.

\section{Acknowledgments}
This work was supported by the Interdisciplinary Thematic Institute SysChem via the IdEx Unistra (ANR-10-IDEX-0002) within the program Investissement d’Avenir.
 
\phantomsection
\bibliography{biblio}
 
\end{document}